
%

\magnification = 1200
\baselineskip = 14 pt

\noindent {\bf The Interpretation of Quantum Mechanics.}
By Roland Omn\`es.  Princeton University Press, Princeton, New Jersey, 1994,
xiv~+~550~pp., \$95.00 (hardback), \$39.50 (paperback).
\vskip 1 truecm

	This monograph is the first book-length treatment of the consistent
histories approach to the interpretation of quantum mechanics which I initiated
in 1984, and to which Omn\`es (starting in 1987) and Gell-Mann and Hartle
(starting in 1990) have made major contributions.  While consistent historians
do not agree on every detail, there is a common core of ideas which can be
summarized as follows. A closed quantum system (the universe, if one is
ambitious) is represented by a Hilbert space, and anything that can sensibly be
said about it at a particular time is represented by some subspace of this
Hilbert space; in other words, there are no hidden variables.  A {\it history}
consists of a sequence of subspaces $E_1, E_2,\ldots$ associated with times
$t_1, t_2,\ldots$, understood as events occurring, or properties which
are true,
at these times.  Provided the history is a member of a {\it consistent family}
of histories, it can be assigned a probability, and within a given consistent
family these probabilities function in the same way as those of a classical
stochastic theory (imagine a sequence of coin tosses): one and only one of them
occurs, and the theory assigns a probability to each possibility.  Inconsistent
histories, those which do not belong to a consistent family, are meaningless.
The unitary time evolution generated by the Schr\"odinger equation, without any
stochastic or nonlinear modifications, is used both for determining the {\it
consistency conditions} which define consistent families, and for calculating
the probabilities of histories belonging to a particular family.

	Measurements play no fundamental role in the consistent history
interpretation; they simply correspond to sequences of events inside a closed
system in which the measurement apparatus, along with everything else, is
treated quantum mechanically.  Thus a possible history for a closed system in
which there is an apparatus for measuring the z component of the spin of a
particle might include an initial state, a value of $S_z$ at a time shortly
before the particle reaches the apparatus, and the position of a pointer on the
apparatus at some later time.  Using conditional probabilities one can show,
under suitable conditions, that the particle earlier had the property indicated
later by the pointer position.  In this and various other ways the consistent
history approach replaces the smoky dragons which inhabit textbook and other
treatments of ``measurement'' with precise mathematical and logical rules
yielding results which are often much closer to the intuition of experimental
physicists than to what one finds in the literature on quantum foundations.

	The first two chapters of the book contain an introduction to quantum
mechanics and the Copenhagen interpretation, thus setting the stage for the
material which follows.  Chapters 3 and 4 give a number of basic tools of the
consistent history approach, including definitions of properties, histories,
and the consistency conditions. Omn\`es' consistency condition resembles my
original proposal[1], and he makes no use of the alternatives suggested by
Gell-Mann and Hartle[2], one of which I myself now prefer.  His generalization
of my definition of a (possibly consistent) family of histories is probably a
step in the right direction, but its physical significance is hard to judge,
since (despite assertions to the contrary on pp. 136, 189, and footnote 19 of
p.  462) all of the examples in the book seem to fit my original proposal.

	Chapter 5 is devoted to the logical framework of the theory, an area in
which Omn\`es has made a major contribution.  The essential idea can be
illustrated in classical terms by imagining tossing a coin three times in
succession.  The result will be one of a set $S$ of 8 possible histories, or
sequences of heads and tails, and a proposition such as ``heads occurred once
on the first two tosses'' corresponds to a subset of $S$, those histories for
which it is true.  Negation of a proposition and the conjunction of two
propositions then correspond to operations on subsets, and logical implication
to one subset falling inside another.  In the quantum case one uses the same
approach, with $S$ the set of elementary histories in a {\it single} consistent
family.  Omn\`es actually sets up the idea of implication using conditional
probabilities rather than set theoretic inclusion, which does not change the
main idea.  (Unfortunately his system as stated does not satisfy the rules of
App.~A to Ch.~5, though this can probably be remedied fairly simply.)  Omn\`es
refers to the propositions, etc. of a single consistent family as a ``logic''.
What distinguishes the quantum from the classical world is that in the former
there are many ``logics'' (consistent families) which can, at least
potentially, refer to the same physical system, but which are mutually
incompatible in the sense that if history $H$ belongs to ``logic'' $L$ and
history $H'$ to ``logic'' $L'$, there may be no ``logic'' which contains both
$H$ and $H'$.  Sound quantum reasoning must conform to Omn\`es' Rule 4 (p.
163): {\it Any description of the properties of an isolated physical system
must consist of propositions belonging together to a common consistent logic.
Any reasoning to be drawn from the consideration of these properties should be
the result of a valid implication or of a chain of implications in this common
logic.}

	I agree with Omn\`es that one must have a rule of this form, and I
think he has expressed it quite well.  But I think still more is needed in
order to specify what constitutes sound reasoning in the quantum domain. In
particular, Rule 4 does not tell us {\it how} to choose the single ``logic''
(consistent family) to use for a reasoning process.  While this choice is
sometimes obvious, in other cases it is not, especially since a single history
can belong to a number of mutually incompatible ``logics''.  This problem
deserves a more careful discussion than it receives in this book or elsewhere
in the consistent histories literature.  It should be noted that the logic of
the consistent history approach is significantly different from that proposed
by Birkhoff and von Neumann[3].

	Omn\`es' formulation differs from my original proposal in the presence
of a density matrix in the consistency condition and probability formulas,
coupled with an interpretation which is not symmetrical under changing the
direction of time. My current ideas[4] escape, I believe, the criticisms in
App.~D of Ch.~5 while still maintaining a structure invariant when reversing
time, and without a density matrix.  I think this shows that Omn\`es'
time-asymmetric version is not a logical necessity, but instead reflects his
use of a time-asymmetric set of axioms.  I should add that, despite certain
differences, there is a great deal of overlap between Omn\`es' ideas and mine,
and their application to various gedanken experiments often leads to identical
consequences.

	Consistent historians all believe that the ``classical world'' of
everyday experience can be understood in quantum mechanical terms by using the
consistent history approach, but differ on the details of how to work out the
correspondence.  Gell-Mann and Hartle [2] suggest finding a consistent family
or families in which classical laws represent an asymptotic approximation to
fully quantum behavior.  Omn\`es' approach, in Ch. 6, is to argue that a
quantum description using quasi projectors corresponding to a cell in classical
phase space can, under suitable conditions, be shown to yield, within errors
which can be estimated, the same time development as the corresponding
classical equations of motion.  As I am not familiar with the mathematical
tools he employs, I cannot comment on the technical aspects of the argument.
Assuming it is correct, it seems to me it is not complete, for the
correspondence holds only for a limited period of time, and in the case of a
system which shows classical chaos, this time can be quite short.  Omn\`es is
not clear about what happens for longer times. He would probably agree with
Gell-Mann and Hartle that some sort of stochastic behavior represents an
appropriate quantum correction to classical laws.  I find the conceptual
foundation of the Gell-Mann and Hartle program a bit clearer, but they, too,
have not worked out all the details.  Both approaches would be impossible
without the freedom, present in the consistent history framework, to talk about
events occurring at successive times, rather than just measurements.

	Chapters 7 considers decoherence, the effect upon a quantum system of
its environment.  Decoherence is an important physical phenomenon, and Omn\`es'
discussion has much of value. I cannot, however, agree with his claim that
decoherence somehow resolves the problem of macroscopic quantum superposition
(MQS) states which besets traditional approaches to a quantum theory of
measurements.  Omn\`es is aware of Bell's criticism that arguments of the ``for
all practical purposes'' type do not resolve fundamental questions, and he
responds that the sort of measurements which could be used to detect the
coherence in MQS states are ``in principle'' impossible, because, for example,
they would require apparatus of physical size larger than the visible universe.
However, it seems to me that the real import of Bell's critique is that if we
understood quantum theory properly, there would be no need to extricate
ourselves from conceptual difficulties by appealing to facts about the world
which are not (or not obviously) part of quantum theory itself, so I do not
find Omn\`es' solution satisfactory.  But in addition there is an alternative
approach within the consistent history scheme itself which, it seems to me,
adequately meets Bell's criticism: one uses a consistent family containing the
outcomes of the measuring process (``pointer positions''), and if one employs
this family or ``logic'', all references to the MQS states are excluded by
Omn\`es' Rule 4: adding them to the discussion would render the family
inconsistent.  Perhaps Omn\`es does not agree with this central result of [1];
it seems not to occur in his discussions of decoherence and measurement.

	Chapter 8, on the theory of measurement, begins with a a derivation of
a fundamental results of the consistent history approach: from the outcome of a
measurement (e.g., the position of a pointer) one can, under suitable
conditions, infer a property of the measured system (e.g., some component of
the spin) at an earlier time.  In other words, quantum measurements (when
properly carried out) actually do what they are supposed to do.  Next comes a
discussion of ``actual facts'', the problem of why in the world around us we
observe only one of the possible situations allowed by quantum theory.  My own
point of view[1] is that quantum theory is irreducibly stochastic, and thus the
occurrence of a particular history is no more (or less) mysterious than the
fact that if a coin is tossed three times in succession, only one of the eight
possible results will actually occur.  Omn\`es evidently does not accept this
point of view. I think he should nonetheless have mentioned it among the
options he lists on p. 493, as it is probably common among working physicists.
The third topic in Ch. 8 is Omn\`es' concepts of ``true'' and ``reliable''.
There are some serious flaws here, as pointed out by Dowker and Kent[5].
Omn\`es[6] believes they can be corrected, but in the meantime, their existence
puts some of his conclusions in doubt.  My own belief is that a more
satisfactory approach would be to associate a separate notion of ``true'' with
each ``logic'', but this idea[7] has not yet been spelled out in detail.

	The discussion of the EPR paradox in Ch. 9 makes use of the ``true''
vs. ``reliable'' distinction just mentioned, and has other problems as well.
What Omn\`es calls the ``original EPR experiment'' is not what one finds in the
original EPR paper, though the paradoxical aspects are closely related.  More
serious, in my opinion, is that Omn\`es makes what one can say about a
particular system dependent on the time at which a measurement is carried out
on a distant system.  This rather unnatural effect occurs in standard quantum
treatments through the ``collapse of the wave function'' (something both
Omn\`es and I reject, at least as a physical phenomenon), and I showed in
1987[8] that it is absent from a consistent history analysis.  Perhaps this is
one of the points at which Omn\`es' use of a time-asymmetric formulation makes
some difference.

	Chapters 10 and 11 of the book contain various applications to physical
systems.  In some instances, as in the decay of unstable particles, Omn\`es'
treatment contains valuable insights.  In others, as in his discussion of
Josephson junctions, it is not clear to me that the discussion adds much to
what is already in the literature on the subject.  Chapter 12 is in two parts:
the first is a summary of material in earlier chapters, and the second treats
various philosophical issues.

	Generally speaking, the book is not easy to read.  The style tends to
be detailed and ponderous, and the overall direction of a particular argument
is often not clear.  Introductory passages where one might have hoped to find
an overview preceding a detailed discussion tend not to be very helpful, for
they contain a large number of peripheral comments and references to other
sections, and occasionally the grammatical construction leaves the meaning of a
sentence unclear.  Sometimes material on a particular topic is divided into two
parts placed in separate chapters for no apparent reason.  The index is helpful
but incomplete.  I am afraid it will not be easy for readers unfamiliar with
the consistent history approach to learn it from this book. On the other hand,
it contains an enormous amount of relevant material, and thus can serve as a
useful reference.

	While it does have defects, I nevertheless admire this book as a bold
attempt to give substance to a vision common to consistent historians, namely
that our current scientific understanding of the physical world, macroscopic
phenomena as well as microscopic, can be linked to a firm foundation of quantum
mechanical principles by appropriate and precise rules of sound reasoning.
There is no need to add hidden variables to the Hilbert space, or tinker with
the Schr\"odinger equation, or restrict ourselves to talk about
``measurements'' in order to construct a coherent interpretation of quantum
mechanics which overcomes the well-known conceptual problems which have given
such trouble to those who have been attempting to understand the subject for
the past seventy years.  We are indebted to Omn\`es for showing how much of
this program can actually be carried out within the consistent histories
framework. Some imperfections are inevitable in a pioneering effort, and those
portions of the book which I find problematical are nonetheless useful as
indications of what needs to be better understood, or more clearly explained,
or perhaps both.

\vskip .3truecm

{\obeylines \parindent 10 truecm
Robert B. Griffiths
Physics Department
Carnegie-Mellon University
Pittsburgh, Pennsylvania 15213
	}

\vskip .5 truecm
{\obeylines \parindent 0 truecm
[1] R. B. Griffiths, J. Stat. Phys. 36 (1984) 219.
[2] M. Gell-Mann and J. B. Hartle, Phys. Rev. D 47 (1993) 3345.
[3] G. Birkhoff and J. von Neumann, Annals of Math. 37 (1936) 823.
[4] R. B. Griffiths, Phys. Rev. Lett. 70 (1993) 2201.
[5] F. Dowker and A. Kent, preprint (1994).
[6] R. Omn\`es, private communication.
[7] R. B. Griffiths, Found. Phys. 23 (1993) 1601.
[8] R. B. Griffiths, Am. J. Phys. 55 (1987) 11.
	}
\end